\documentstyle[PASJadd, epsf]{PASJ95}

\begin{document}

\title{Structural Model of Molecular Cloud Complexes: 
Mass, Size, and External Pressure}

\author{Akio K. {\sc Inoue} and Hideyuki {\sc Kamaya} \\
{\it Department of Astronomy, Faculty of Science, Kyoto University,
Sakyo-ku, Kyoto 606-8502, JAPAN} \\
{\it E-mail(AKI): inoue@kusastro.kyoto-u.ac.jp}}

\abst{
We investigate the structure of the molecular cloud complexes (MCCs) as
a group of several giant molecular clouds (GMCs) in the Galaxy.
Then, we find that the mass--size relation which has been reported for
the GMCs establishes well even for the very large MCCs whose size is
about 1 kpc.
Since the horizontal size of the MCCs is more than the
thickness of the Galactic disk, we can no longer consider the MCCs to be
a sphere.
Thus, we construct a structural model of the MCCs, adopting a
rectangular-solid geometry.
As a result, our model explains the observed mass--size relation of the
MCCs very well.
{}From the estimated external pressure around the MCCs, we find that
they are in a rough pressure balance with the interstellar medium.
Moreover, we find there is observational deficiency of the MCCs with a
large size and surface density.
Then, we suggest that the external pressure has a significant effect on 
the structure and evolution of the MCCs.
We also discuss the effect of H {\sc ii} regions in the MCCs.
}
 
\kword{hydrodynamics --- ISM : clouds --- 
ISM : structure ---  Galaxy: ISM --- stars : formation}

\maketitle
\thispagestyle{headings}

\section{Introduction}

The star forming regions of the galaxies have a kind of the hierarchical
structure; individual stars, H {\sc ii} regions, OB associations,
aggregates, giant molecular clouds (GMCs), and complexes (Elmegreen,
Salzer 1999).
The star-forming complexes are the regions containing the smaller
structures, and their typical size is more than several hundreds parsecs.
For example, in our Galaxy, there are star complexes whose averaged
diameter is about 600 pc (Efremov 1978).
Also, there are H {\sc i} superclouds which include a group of the GMCs
(Elmegreen, Elmegreen 1987).
In M51, the existence of the giant molecular associations, whose mass is
more than $10^7 \MO$, are reported by Vogel, Kulkarni, Scoville (1988)
and Rand, Kulkarni (1990).
In this $Letter$, we concern such star-forming complexes, especially,
the molecular cloud complexes (MCCs) as a group of several GMCs in order 
to understand the star formation activity over a galactic wide scale.
Such MCCs have been observed in our Galaxy (Myers et al.\ 1986),
although they are smaller than the giant molecular associations in M51.

By the way, from many observations of the GMCs, it has
reported that there are famous correlations; the size--velocity
dispersion relation and the size--mass (or density) relation (e.g.,
Larson 1981; Myers 1983; Sanders et al.\ 1985; Dame et al.\ 1986;
Solomon et al.\ 1987).
The former relation implies that the interstellar medium (ISM) is in a
turbulent condition.
The latter is that the mass of the GMCs is proportional to the square
of their size.
However, for the MCCs, these empirical correlations mentioned above
have not discussed sufficiently to date.
Hence, we discuss the mass-size relation (and also surface density-size
relation) for the MCCs in this $Letter$.

On the other hand, the famous observational paper by Myers (1978) shows
clearly that there is a global pressure balance among the variety
components of the ISM.
Although Bowyer et al.~(1995) shows an evidence for a pressure imbalance
in the local ($\ltsim$ 40 pc) ISM, such a pressure balance can be still
acceptable in a galactic scale.
Thus, a kind of the pressure equilibrium is often adopted to examine
the structure and evolution of any component of the ISM.
Indeed, the criterion of the gravitational contraction of the
self-gravitating clouds is affected by the external pressure around
them (e.g., Ebert 1955; Bonnor 1956; Nakano 1998).
Thus, we should always pay a part of our attention to the importance of
this external pressure.

Assuming such an pressure equilibrium and a sherical geometry, the
structure and the empirical correlations among mass, size, and velosity
dispersion of the GMCs are studied very well (e.g., Chi\`{e}ze 1987;
Maloney 1988; Elmegreen 1989; Mckee, Holliman 1999).
The assumption of the spherical structure for
each GMC is not so crucial statistically 
if the number of sample clouds is sufficient.
However, it is not good for the MCCs.
Indeed, the MCCs are not spherical (see, for
example, Figure~3 of Myers et al.\ 1986), and
their horizontal size is about
or above the thickness of the disk of the host galaxies.
In this $Letter$, therefore, we investigate the structure of the MCCs,
adopting not spherical geometry model.

In the next section, 
we summarize the property of the data adopted in our discussions.
Our model is described in section 3.
The results and discussions are presented in section 4, 
and the summary is presented in the final section.

\section{Data Properties}

We need a large sample of the MCCs, to
obtain their clear and convincing property as a general picture. 
Furthermore, it is better that the sample MCCs are relatively similar
each other.
Then, we choose the MCCs in the Galaxy as the sample for our
analysis.
The suitable data of the MCCs are compiled in Myers et al.\ (1986).

Myers et al.\ (1986) examined 54 molecular clouds and cloud complexes in the
Galactic disk.
These clouds and complexes locate in $-1^\circ \leq b \leq 1^\circ$,
$12^\circ \leq l \leq 60^\circ$.
The MCCs contain typically five local maxima of the
observed CO intensity in themselves.
Solomon et al.\ (1987) observed the same region and detected 273 GMCs.
These two observations are supplemented each other because of their
different spatial resolution.
For example, Solomon's No.45, 48, 49, and 56 objects seem to correspond
to Myers' 17,58 objects.
Also, Myers et al.\ (1986) examine whether individual clouds and MCCs
are associated with H {\sc ii} regions.
Some properties of their sample are tabulated in their Table~2.

The mass of each MCC was estimated from the
integrated CO intensity over its observed area by using a standard
CO--H$_2$ conversion factor ($2\times10^{20}$ cm$^{-2}$ [K km
s$^{-2}$]$^{-1}$ of Lebrun et al.\ 1983).
The derived observational masses distribute over the range of
$10^{4-7}\, \MO$, and the median value is $6.3\times 10^5 \MO$.

We can observe the projected size of the MCCs
perpendicular to the line of sight.
In this $Letter$, we define a horizontal size of each MCC, $l$, by the
following equation,
\begin{equation}
 l=D\tan\delta l\,\,[{\rm pc}]\,,
\end{equation}
where $D$ denotes the distance in kpc from the Sun to the MCCs given by
the column~(4) of Table~2 in Myers et al.\ (1986) and $\delta l$ is
defined in degree by $l_{\rm max}-l_{\rm min}$, which are the maximum
and minimum Galactic longitude of the location of samples also given by
the columns~(2) and (3), respectively.
One of the sample clouds of Myers et al.\ (1986) has a very small size
($l=7.9$ pc). 
Since we focus on the MCCs, we exclude it from the sample in
our analysis.
The determined $l$ of the sample MCCs distributes over
the range from 40 pc to more than 1 kpc.
The mean value of $l$ is about 300 pc.
Here, we must note that the vertical size of the sample complexes is not 
available from Myers et al.\ (1986).
Then, it is determined via a structural model presented in section 3.

A typical cloud complex has $6\times10^5 \MO$ as its mass and 300 pc
as its size.
Then, the escape velocity against its gravitational potential is about
3 km/s.
This is almost same with the typical velocity dispersion of the internal 
GMCs (Solomon et al.\ 1987) and the stellar populations in the Galactic
disk.
Thus, we cannot insist that the MCCs are self-gravitating, and the MCCs
may be the coincident aggregates of the GMCs.
Indeed, the MCCs in the inter-arm regions may be unbound, but it is shown 
that the superclouds in the arm regions cannot be reproduced by a simple
random superposition of the GMCs (Rand, Kulkarni 1990).
Moreover, our sample MCCs are the internal structure of the H {\sc i}
superclouds which are approximately vilialized objects (Elmegreen,
Elmegreen 1987).
Therefore, we assume the MCCs to be bounded objects.

In this sample of the MCCs, there is a good correlation between
the mass estimated from the integrated CO intensity and the size of the
perpendicular to the line of sight, that is, their masses are
proportional to the square of their sizes.
This is shown in figure~1.
This relation is the same with that of the GMCs, although the both
relations differ in the range of size.
That is, the well known mass--size relation of the GMCs is established
up to the MCCs whose size is about 1 kpc.

Moreover, according to Elmegreen, Salzer (1999), the blue luminosity of
the star-forming complexes in spiral and irregular galaxies, whose mass
and size are equivalent to that of the MCCs of this $Letter$, is also
proportional to the square of their size.
If we consider the blue luminosity to be proportional to the mass of the
region, the result of Elmegreen, Salzer (1999) is consistent with our
figure~1.
Thus, the mass--size relation may be universal over large dynamic
range of the size and mass.

\section{Model Description}

In this $Letter$, we focus on the MCCs in the Galaxy.
As mentioned in section 2, a typical size along the Galactic plane of the
sample complexes is more than 100 pc which is the thickness of the
Galactic disk. 
Thus, the horizontal size of the MCCs is too large for us to consider
the MCCs to be a sphere, unfortunately.
In addition, the mechanism to determine a typical horizontal size of the
MCCs should be different from that of the vertical direction.
That is, the shear of the Galactic rotation must affect the
determination of the horizontal size because 
of their large ``Galactic scale'' size, while the vertical scale is little
affected by the rotation.
Therefore, we assume a rectangular-solid geometry for the MCCs.
This is the most simple geometry, except for the spherical one.

Let us consider below 
the mass of the MCCs with the rectangular-solid geometry.
Since the physical length of the MCCs parallel to the line of
sight is never observed in the Galaxy, this length is assumed to be the
same with the size of perpendicular to the line of sight, $l$, defined
in section 2.
First, using this $l$, we define the expected mass of a MCC as 
\begin{equation}
 M = \rho_{\rm m} h l^2 \,,
 \label{modelmass1}
\end{equation}
where $\rho_{\rm m}$ and $h$ are a mean density and a vertical thickness of
a MCC, respectively.

Next, we discuss the vertical scale of the MCCs.
The Galactic rotation dose not affect the determination of the size in
the vertical direction.
Thus, we estimate that the vertical size of the MCCs is the simple Jeans
length.
That is, $h=\sqrt{\pi {c_{s,{\rm eff}}}^2/G\rho}$, where
$c_{s,{\rm eff}}$ represents an effective sound speed in a complex which
contains the effects of thermal motions, turbulence, and magnetic
fields, and $\rho$ is a mean density of the Galactic disk.
If we represent the mean stellar density as $\rho_*$, then, $\rho =
\rho_{\rm m} + \rho_*$.
$c_{s,{\rm eff}}$ is given by $\sqrt{p_{\rm eff}/\rho}$, where
$p_{\rm eff}$ denotes an internal effective pressure inside a MCC
(e.g., Kamaya, Shchekinov 1998; Kamaya 1999).
In this $Letter$, we approximate $p_{\rm eff}= a p_{\rm ex}$ ($a>1$),
where $p_{\rm ex}$ is an external pressure and $a$ is a factor of order
of unity, because the internal pressure connects to the external one
continuously, so both pressures are same order.
Also, we approximate $\rho = b \rho_{\rm m}$, i.e., $\rho_* = (b-1)
\rho_{\rm m}$ ($b>1$).
Since $\rho_* \sim 0.1-1 \MO {\rm pc}^{-3}$ (Binney, Merrifield 1998),
and $\rho_{\rm m} \sim 0.7 \MO {\rm pc}^{-3}$ for the typical MCC
(where we remember the scale hight, $h$, is about 100 pc), we
consider $b\sim 1-2$.
Therefore, we obtain the vertical thickness of a MCC as the
following equation;
\begin{equation}
 h = \frac{{p_{\rm ex}}^{1/2}}{\rho_{\rm m}}\sqrt{\frac{\pi}{G}}\,,
 \label{thickness}
\end{equation}
where we set $\sqrt{a}/b \sim 1$.
{}From this equation, we find that the vertical thickness multiplied by
the density of the MCCs, that is, $h \rho_{\rm m}$ which is considered
to be a kind of the face-on surface density, depends on only the
external pressure, $p_{\rm ex}$ (see also Inoue et al.\ 2000).

Finally, we eliminate $\rho_{\rm m}$ and $h$ from equation~(\ref{modelmass1})
by equation~(\ref{thickness}), then, we obtain the expected mass of the
MCCs;
\begin{equation}
 \log \left(\frac{M}{\MO}\right)
  = 2\log \left(\frac{l}{\rm pc}\right)
   + \frac{1}{2}\log\left(\frac{p_{\rm ex}}{\rm k_{\rm B}\,K\,cm^{-3}}\right) 
   - 0.413 \,,
 \label{modelmass2}
\end{equation}
where $k_{\rm B}$ is the Boltzmann's constant.
We compare this expected mass with the observational mass of the
MCCs in the next section.

\section{Results and Discussions}

\subsection{Mass vs Size}

In figure~1, we compare the observational mass 
of the sample of Myers et al.\ (1986) (excluded one very small cloud) 
with the theoretical lines calculated by utilizing
equation~(\ref{modelmass2}).
The filled points represent the observed data of the MCCs
associating with H {\sc ii} regions and the open points are
those of the MCCs without H {\sc ii} regions.
The solid, dotted, dashed, and dash-dotted lines are calculated 
via equation~(\ref{modelmass2}) by adopting
$p_{ex}/k_{\rm B}=10^2$, $10^3$, $10^4$, and $10^5$ K cm$^{-3}$,
respectively.

In this figure, we find that our model lines reproduce the observational 
correlation of the mass with the size of the MCCs.
We also find from this figure that the external pressure around the cloud
complexes is expected to be about $10^{2-4}\,{\rm k_B\,K\,cm^{-3}}$.
The average value is 
$\langle\log(p_{ex}/{\rm k_B\,K\,cm^{-3}})\rangle=3.4\pm0.7$, 
where we also show the standard-deviation.
This pressure is consistent with the pressure roughly balancing among
various components of the ISM reported by Myers (1978).
This indicates that the MCCs are in an equilibrium state along
the rough pressure balance of Myers (1978), while the GMCs, whose size
is about 10--50 pc, are not generally in such an equilibrium.

The average pressure of the MCCs estimated here is less than the
turbulent pressure of their parent H {\sc i} clouds determined by
Elmegreen, Elmegreen (1987).
This can be understood naturally if we consider that the regions with
low turbulent motion in the parent H {\sc i} clouds evolve into the
molecular clouds.
In fact, the cold and quiescent (i.e., low velocity dispersion or narrow 
emission line width) H {\sc i} clouds in dwarf irregulars associate
with the star-forming regions (Young, Lo 1996).
Since the star-forming regions are considered to be within the molecular
clouds, it indicates that the turbulent motion around the molecular
clouds are relatively low.
Therefore, such low external pressure of the MCCs is reasonable.

Finally, we comment on the method for estimating the mass of the MCCs or
the star-forming regions.
Once an external pressure is given, we can determine the mass of the
MCCs from the observation of their size, by using
equation~(\ref{modelmass2}).
Moreover, if a MCC is the parent cloud of a star-forming region, and the
sizes of the both objects are nearly the same, we can determine the mass
of the star-forming region from only its size and a proper external
pressure.
Therefore, equation~(\ref{modelmass2}) may be very useful.

\subsection{Face-On Surface Density vs Size relation}

To discuss the relation between the surface density and size, we define
a face-on surface density, $\Sigma$, by the observational mass and size
of the sample MCCs as being $\Sigma \equiv M/l^2$.
In figure~2, we show the relation between this surface density and the
size.
The lines are calculated by the following equation;
\begin{equation}
 \log \left(\frac{\Sigma}{\MO {\rm pc}^{-2} }\right)
  = \frac{1}{2} \log \left(\frac{p_{ex}}{\rm k_{\rm B}\,K\,cm^{-3}}\right) 
   - 0.413 \,,
 \label{SD}
\end{equation}
which is derived from equation~(\ref{thickness}).
The filled and open points stand for the same mean as figure~1.

We find evidently the void region of observational points at the region
of a large surface density and a large size in figure~2.
In other words, the large MCCs are not observed under a high
pressure condition along the context of our rectangular-solid model.
If the deficiency is real, we suggest that a higher external pressure
makes the size of the MCCs smaller.
It may mean that the MCCs in a high pressure evolve efficiently.
Indeed, a higher external pressure is likely to compress the MCCs more
easily than a lower pressure.
Then, cloud components inside each MCC may collide each other and
the MCCs evolve rather quickly.
Inversely, under the condition of a low external pressure, 
even the large MCCs can survive and evolve slowly.
Thus, we conclude that the external pressure is important to determine
the structure of the MCCs.

Unfortunately, we must comment that the uncertainty of the size
determination of sample MCCs is large. 
In fact, Myers et al.\ (1986) includes the uncertainty of a factor of $\sim 2$
for their determination of the cloud's boundary.
Then, we must note that our result also includes this observational
uncertainty.
Hence, it is possible to think that the upper-right void is a just
observational bias.

\subsection{Effect of H {\sc ii} Regions}

Since H {\sc ii} regions may affect the structure of the
MCCs, we divide the sample into two
groups in order to examine this effect.
One is the group of the MCCs associating with observed H
{\sc ii} regions, and the other is that of the MCCs without
these regions.
In figures~1 and 2, we find the comparison between this divided data
points and model lines.

If there are H {\sc ii} regions in the MCCs, the
effective pressure in the MCCs should be larger than that of the
MCCs without H {\sc ii} regions. 
This means that the suitable external pressure is rather high for the
MCCs with H {\sc ii} regions. 
In fact, for the MCCs with H {\sc ii} regions,
$\langle\log(p_{ex}/{\rm k_B\,K\,cm^{-3}})\rangle=3.5\pm0.8$, while
for the sample without H {\sc ii} regions,
$\langle\log(p_{ex}/{\rm k_B\,K\,cm^{-3}})\rangle=3.1\pm0.7$, where we
also show the standard-deviation.
Thus, we find this trend in figures~1 and 2, although it is rather weak.
This indicates that the effect of H {\sc ii} regions on the entire
structure of the MCCs dose not so significant.
However, for the individual GMCs in each MCC, the effect may be
critical.
It is inportant to examine whether the pressure of the GMCs with H {\sc ii}
regions is higher than those without the regions.
Unfortunately, we cannot resolve the problem in the framework of this
$Letter$.
Thus, it is an interesting future work.

\section{Summary}

We examine the mass--size relation and the structure of the MCCs in the
Galaxy.
Here we summarize our findings.
First, we find that the mass--size relation which has been reported for the
GMCs establishes well even for the very large MCCs.
This relation is reproduced by our rectangular-solid model of the
strucure of the MCCs.
In our model, the typical external pressure of the MCCs
is estimated to be $2.5\times10^3\,{\rm k_B\,cm^{-3}\,K}$.
Thus, the MCCs may be in raugh pressure balance of Myers (1978).
Also, we find the observational deficiency of the MCCs with a large size
and a large face-on surface density.
And, we find the existence of H {\sc ii} regions in the MCCs increases the
expected pressure slightly.
\par

\vspace{1pc}\par

We would like to thank T. Kawano for his useful suggestion and
discussion, and also to thank M. Sait\={o} for
continuous encouragement.

\section*{References}
% \small

\re
 Binney, J., Merrifield, M.\ 1998, Galactic Astronomy (New Jersey:
 Princeton University Press) ch3, 10

\re
 Bonnor, W.B.\ 1956, MNRAS 116, 351

\re
 Bowyer, S., Lieu, R., Sidher, S.D., Lampton M., Knude J.\ 
 1995, NATURE, 375, 212

\re
 Chi\`{e}ze, J.P.\ 1987, A\&A 171, 225

\re
 Dame, T.M., Elmegreen, B.G., Cohen, R.S., Thaddeus, P.\ 1986, ApJ 305, 892

\re
 Ebert, R.\ 1955, Z.\ Astrophys. 37, 216

\re
 Efremov, Yu.A.\ 1978, Sov.\ Astron.\ Lett. 4, 66

\re
 Elmegreen, B.G.\ 1989, ApJ 338, 178

\re
 Elmegreen, B.G., Elmegreen, D.M.\ 1987, ApJ 320, 182

\re
 Elmegreen, D.M., Salzer, J.J.\ 1999, AJ 117, 764

\re
 Inoue, A.K., Hirashita, H., Kamaya, H.\ 2000, AJ in press

\re
 Kamaya, H.\ 1999, PASJ 51, 617

\re
 Kamaya, H, Shchekinov, Yu.A.\ 1998, PASJ 50, 621

\re
 Larson, R.B.\ 1981, MNRAS 194, 809

\re
 Lebrun, F., Bennett, K., Bignami, G.F., Caraveo, P.A., Bloemen,
 J.B.G.M., Hermsen, W., Buccheri, R., Gottwald, M., Kanbach, G.,
 Mayer-Hasselwander, H.A.\ 1983, ApJ 274, 231

\re
 Maloney, P.\ 1988, ApJ 334, 761

\re
 McKee, C.F., Holliman II, J.H.\ 1999, ApJ 522, 313

\re
 Myers, P.C.\ 1978, ApJ 225, 380

\re
 Myers, P.C.\ 1983, ApJ 270, 105

\re
 Myers, P.C., Dame, T.M., Thaddeus, P., Cohen, R.S., Silverberg, R.F.,
 Dwek, E., Hauser, M.G.,\ 1986, ApJ 301, 398

\re
 Nakano, T.\ 1998, ApJ 494, 587

\re
 Rand, R.J., Kulkarni, S.R.\ 1990, ApJ 349, L43

\re
 Sanders, D.B., Scoville, N.Z., Solomon, P.M.\  1985, ApJ 289, 373

\re
 Solomon, P.M., Rivolo, A.R., Barrett, J.,  Yahil, A.  1987, ApJ 319, 730

\re
 Vogel, S.N., Kulkarni, S.R., Scoville, N.Z.\ 1988, nature 334, 402

\re
 Young, L.M., Lo, K.Y.\ 1996, ApJ 462, 203

\clearpage

\centerline{Figure Captions}

\bigskip

\begin{fv}{1}
{1cm}
{Comparison between the observational mass of the Galactic
MCCs and the lines expected theoretically. 
The points denote the observational data of the sample in Myers et al.\ (1986)
excluded one very small cloud.
The filled points denote the MCCs associated with H
{\sc ii} regions and the open points are without these
regions.
Each of the lines are calculated by equation~(\ref{modelmass2}). 
The solid, dotted, dashed, and dash-dotted lines correspond to
$p_{ex}/k_{\rm B}=10^2$, $10^3$, $10^4$, and $10^5$ K cm$^{-3}$,
respectively.}
\end{fv}

\begin{fv}{2}
{1cm}
{The estimated face-on surface density vs the observational
size for the Galactic MCCs.
The filled and open points mean the same as figure~1.
The lines which are also the same as figure~1
are calculated by equation~(\ref{SD}).}
\end{fv}

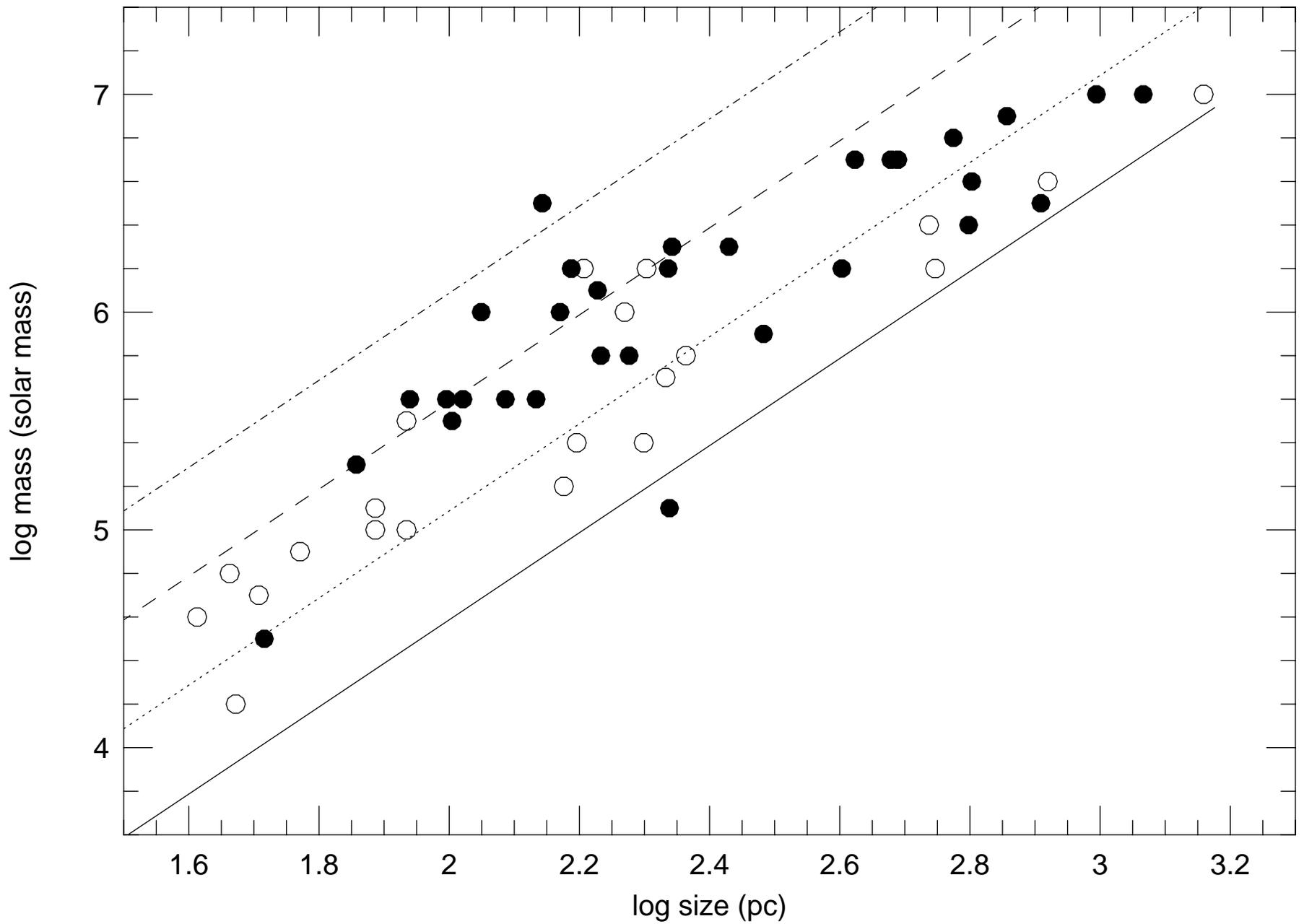
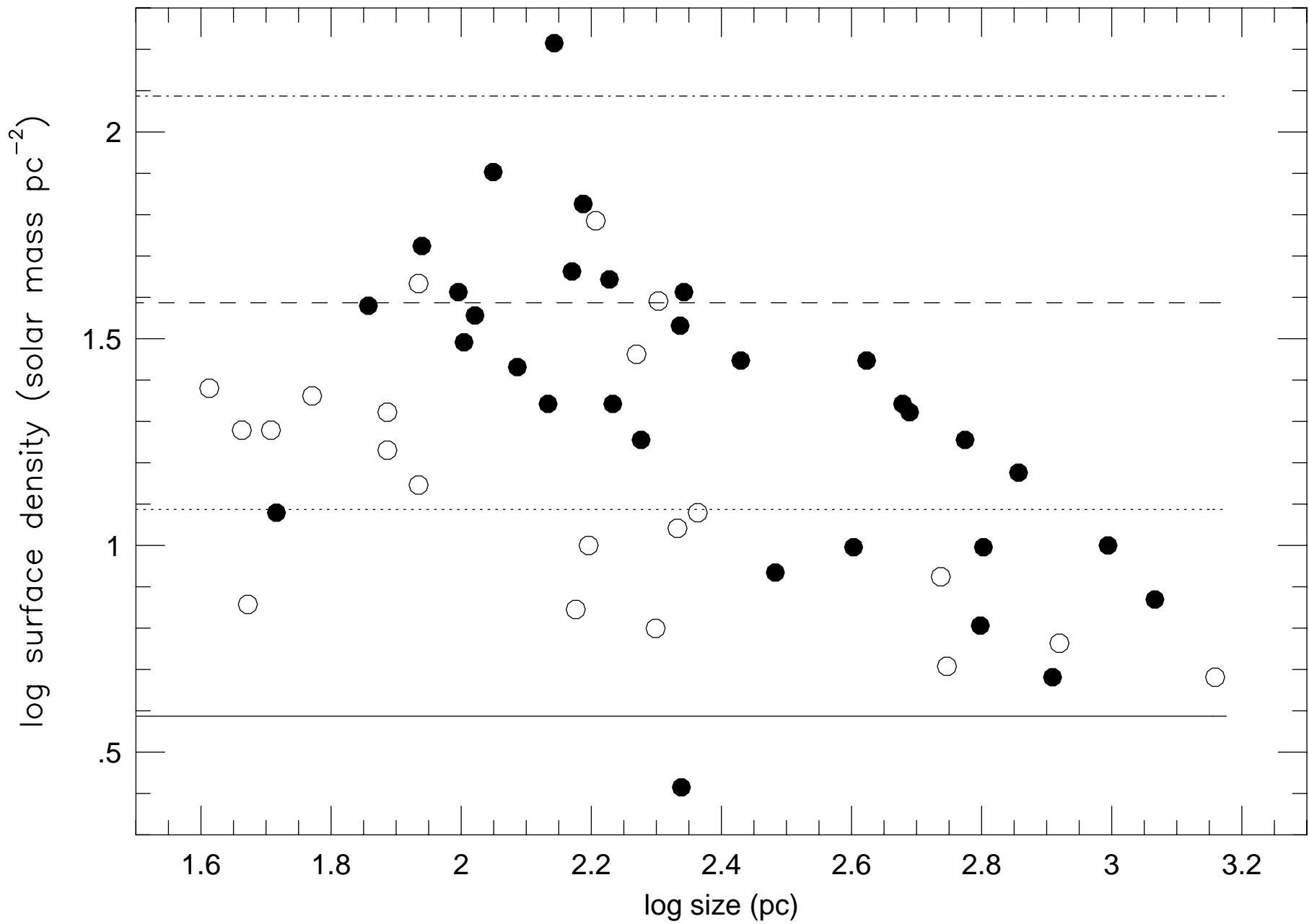
\end{document}